\documentclass[twocolumn,prb,showpacs,amsmath,floatfix]{revtex4}
\usepackage{graphicx}
\usepackage{times}
\begin{document}

\title{Temperature Dependence of Conductance
and\\ Thermopower Anomalies of Quantum Point Contacts
}

\author{O. A. Tkachenko}\email{oatkach@gmail.com}
\author{V. A. Tkachenko}

\affiliation{
%A.V.
Rzhanov Institute of Semiconductor Physics SB RAS,
630090 Novosibirsk, Russia
}

\date{\today}

\begin{abstract}
It has been shown within the Landauer single-channel approach that the presence
of the 0.7 anomaly in the conductance of a ballistic microcontact
and the respective plateau in the thermopower
implies unusual pinning of the potential barrier
height $U$ at a depth of $k_BT$ below the Fermi level $E_F$.
A simple way of taking into account the effect of electron-electron
interaction on the profile and temperature dependence of a smooth
one-dimensional potential barrier in the lower spin degeneracy subband
of the microcontact has been proposed. The calculated temperature dependences
of the conductance and Seebeck coefficient agree with the experimental
gate-voltage dependences, including the emergence of anomalous plateaux
with an increase in temperature.
\end{abstract}


\maketitle
\section{Intoduction}
Quantization of the conductance of submicron constriction
in two-dimensional electron gas~\cite{Wharam}
is described well by the Landauer
formula under the assumption of spin degeneracy of
one-dimensional single-particle subbands in zero
magnetic field.\cite{Glazman,Buttiker,pyshkin,liang}
The same approach explains the
alternation of zero plateaux and peaks of thermopower
(Seebeck coefficient S), which obeys the Mott formula,
$S^M\propto\partial\ln G(V_g,T)/\partial E_F$.\cite{Streda,Proetto,Molenkamp,Appleyard98,Lunde}
However, the dependence $G(V_g)$
of the conductance on the gate voltage
exhibits a narrow region of anomalous behavior, the
$0.7\cdot 2e^2/h$ plateau.\cite{Thomas}
This plateau broadens with an increase in
temperature,\cite{Thomas,Thomas98,Kristensen,Cronenwett,Liu,Komijani}
can disappear at $T\to 0$,\cite{Thomas98,Kristensen,Cronenwett,Liu,Komijani}
but persists at a complete thermal
spread of the conductance quantization
steps.\cite{Appleyard,Cronenwett,Liu}
The 0.7 conductance anomaly is closely related to
the anomalous plateau $S\ne0$ of the thermopower,\cite{Appleyard}
which implies violation of the Mott approximation
$S\propto \partial\ln G(V_g,T)/\partial V_g$.\cite{Appleyard98}
There are dozens of works attempting to explain the 0.7 anomaly
(see Refs.~\onlinecite{Sloggett}, \onlinecite{Micolich} and references therein).
Numerous scenarios including spin polarization, Kondo effect,
Wigner crystal, charge-density waves, and the formation of
a quasi-localized state have been suggested.
Calculations that reproduce the unusual temperature
behavior of the 0.7 conductance anomaly have been
performed so far only within phenomenological fitting
models with spin subbands,\cite{Cheianov}
or beyond the Landauer formula.\cite{Sloggett}
Although a particular mechanism
of the appearance of the anomalous plateaux in conductance
and thermopower remains unclear, their
common reason is thought to be the electron-electron
interaction, which should manifest itself most effectively
at the onset of filling the first subband, where the
electron system is one-dimensional.

In this work, we start from the standard Landauer
approach to the description of conductance and
thermopower of a single-mode
ballistic quantum wire with spin degeneracy.
This approach take into account interaction
via the $T$-dependent one-dimensional reflecting barrier.
First, we show that the appearance of the
0.7 anomaly implies pinning of
the barrier height $U$ at a depth of $k_BT$
below the Fermi level $E_F$  and that this
pinning yields the plateau $S\ne0$.
Next, motivated by
description of Friedel oscillations surrounding
a delta-barrier
in a one-dimensional electron gas,\cite{Matveev94}
we suggest a simple formula which reduces the
$T$-dependent part of the correction to the
interaction-induced potential to the temperature
dependence of the one-dimensional electron density.

The behavior of the conductance and Seebeck coefficient
calculated with the corrected potential agrees with
the published experimental
results.

\section{Basic formulas and unusual pinning}
Conductance and Seebeck coefficient of single mode ballistic channel
can be written within the Landauer approach as follows:\cite{Streda,Proetto,Molenkamp}
\begin{equation}\label{eq-1n}
\begin{gathered}
G=\frac{2e^2}{h} \int_{0}^{\infty} D(E,U(x,V_g))F(\epsilon)dE, \\
S=-\frac{2ek_B}{hG}\int_{0}^{\infty}D(E,U(x,V_g))\epsilon F(\epsilon)dE,
\end{gathered}
\end{equation}
where $D$ is the transmission coefficient, $E$
is the energy of ballistic electrons,
$U(x,V_g)$  is the effective $T$-dependent one-dimensional
barrier, $\epsilon=(E-E_F)/k_BT$,
$F(\epsilon)=4k_BT\cosh^{-2}(\epsilon/2)$
is the derivative of the Fermi distribution
function with respect to $-E$.
There is also the Mott approximation generalized for arbitrary
temperatures $T$:\cite{Appleyard,Lunde}
\[
S^M=-\frac{\pi^2k_B^2T}{3e}\frac{\partial\ln G(V_g,T)}{\partial E_F}.
\]

If the barrier $U(x)$ in a one-dimensional channel is
sufficiently wide (according to the three-dimensional
electrostatic calculations of gate-controlled quantum
wires the barrier half-width must be $\stackrel{>}{_{\sim}}200$~nm),
the step in the energy dependence $D(E)$ of the
transmission coefficient is abrupt and the
respective transition is much narrower in the energy $E$
than the thermal energy $k_BT$, at which the 0.7 plateau occurs.
This condition is satisfied in many experiments.\cite{Thomas98,Kristensen,Cronenwett,Liu,Komijani}
Then, Eq.~\ref{eq-1n} for $G$, $S$ and Mott approximation
gives
\begin{equation}\label{eq-2n}
\begin{gathered}
G= \frac{2e^2}{h}(1+e^{-\eta})^{-1},\\
S=-\frac{k_B}{e}[(1+e^{-\eta})\ln (1+e^{\eta})-\eta],\\
S^M=-\frac{k_B}{e}\frac{\pi^2}{3}(1+e^{\eta})^{-1},
\end{gathered}
\end{equation}
where $\eta=(E_F-U)/k_BT$ and $U=U(x=0)$ is the height of
the barrier $U(x,T,V_g)$.
Clearly, the dependences $G(E_F-U)$ at different $T$
are simply smooth steps of unit height with the fixed common point
$G=e^2/h$.
Curves $S^M(\eta)$ and $S(\eta)$ are numerically close to each other
at the interval $0<S<2k_B/e$ (see Appendix \ref{appA}).
The values $G\approx0.7\cdot 2e^2/h$ in $G(\eta)$
are not particularly
interesting except that they correspond to $\eta\approx 1$.
However, in experiments, there appear plateaux of $G(V_g)$ at these
values, which implies pinning of $U(V_g)$ at a depth of $k_BT$
below the Fermi level (see Appendix \ref{appA}).
According to Eq.\eqref{eq-2n}, the discovered pinning can be expected
to give the plateau $S\approx-0.8k_B/e$ ($S^M\approx-k_B/e$)
in the curve $S(V_g)$.
Appendix \ref{appA} compares the calculated plateaux with the experimental
ones,\cite{Appleyard} and shows that parameter $\eta$ obtained
from $G$ is equal to that from $S$, which verifies
applicability of Eq.\eqref{eq-2n} to the experiment.
Notice that this pinning differs from the pinning
discussed earlier\cite{hansen,Cheianov,Kristensen,Appleyard,Zozoulenko,Pepper}
by unusual temperature dependence and single-channel
%mode
transmission.
The pinning that we detected seems paradoxical
and urges us to suggest that a probe ballistic electron at
the center of the barrier would ``see'' the potential
$U(T, V_g)$, which is different from the potential $V(T, V_g)$
computed self-consistently with the electron density (see Appendix \ref{appB}).
In fact, similar to our previous calculations,\cite{pyshkin,liang} we
computed three-dimensional electrostatics of single-mode quantum
wires using different kinds of self-consistency between the potential
and the electron density with\cite{hansen} and without the inclusion
of exchange interaction and correlations in the local approximation.
These calculations show quite definitely that the one-dimensional
electron density $n_c$ in the center of the
barrier is almost independent of $T$ and is linear in $V_g$
starting from small $n_{c0}$ values; i.e., the electric capacitance
between the gate and the quantum wire is conserved at $G > e^2/h$.
In addition, since the density of states is positive, $d(E_F-V)/dn_c > 0$
and the dependence $V(V_g)$ of the self-consistent barrier height on the
gate voltage does not yield pinning even with the
inclusion of the exchange-correlation corrections in
the local approximation (Appendix \ref{appB}). Therefore, we suggest that
the discovered pinning of the reflecting barrier height
is due to the nonlocal interaction.

\section{Estimation of nonlocal interaction}
It is well known in atomic physics and physics of
metallic surfaces and tunneling gaps between two metals
that the potential seen by a probe electron in a low-density
region is different from the self-consistent
potential found with the inclusion of interaction in the
local approximation.\cite{Bardeen,Latter,Binnig,Bertoni}
This difference was
attributed to nonlocal exchange and correlations, i.e., to the
attraction of the electron to an exchange-correlation
hole, which remains in a high-density region.\cite{Gunnarsson,Wang}
Consideration of this phenomenon regarding a quantum
point contact is currently unavailable, despite the
obvious analogy between two metallic bars separated
by a tunneling gap and two-dimensional electron gas
baths separated by a potential barrier. We suggest that
a ballistic electron coming to the barrier region, where
the density $n_c$ is low, gets separated from its
exchange-correlation hole, which is situated in the region of
dense electron gas. As a result, the local description of
the correction to the potential becomes inadequate.
Although the hole has a complicated shape, it can be
associated with the effective center. Then, a decrease
in the potential for the ballistic electron in the center
of the barrier is
$U-V\approx-\gamma e^2/(4\pi\epsilon\epsilon_0 r)$, where
$r$ is the
distance between the centers of the barrier and hole,
whereas $\gamma\stackrel{<}{_{\sim}}1$
takes into account the shape of the hole
and weakly depends on $r$. In perturbation theory, we
are interested in a small (i.e., $T$-dependent) part of the
correction:
\begin{equation}\label{deltaU}
\delta U\approx-[e^2/(4\pi\epsilon\epsilon_0]\gamma(r(T)^{-1}-r(0)^{-1}),
\end{equation}
and the correction at $T=0$ is thought to be already
included in the independent variable, which is the initial
barrier $U_0(x)$.
Obviously, $r$ decreases with an
increase in $n_c$, until the electron and hole recombine
and the local approximation for the interaction term
becomes valid in the center of the barrier.
According to this tendency and the smallness of the $T$-dependent
correction, we can write
$\gamma (r(T)^{-1}-r(0)^{-1})\approx (n_c(T)-n_c(0))/(r^*n_c^*)$, where
$n_c(T)$ and $n_c(0)$ are found perturbatively from the single-particle
wavefunctions in the barrier $U_0(x)$.
In a certain range of the barrier height,
we can also neglect a change in the positive phenomenological parameter
$\gamma r^*n_c^*$. Under these assumptions,
Eq.\eqref{deltaU} is formally a special case of the interaction-induced correction
$\delta U(x)\propto-\alpha\delta n(x)$,
here, $\delta n(x)$ stands for Friedel density
oscillations. This correction results from calculation
of the propagation through the delta barrier in a one-dimensional
electron system,\cite{Matveev94} in which case
$\alpha=\alpha(0)-\alpha(2k_F)$ is a result of the competition between
the exchange ($\alpha(0)$) and direct ($\alpha(2k_F)$) contribution
to the interaction. A similar correction was used to
simulate multimode quantum wires.\cite{Renard} We can
attempt to extend the range of this correction, with the
respective change in the meaning of $\alpha$ to the entire
first subband of the quantum wire, including the top of
the smooth barrier.

\section{Calculation of 1D electron density}
\begin{figure}[t]
\centerline{\includegraphics*[width=0.98\linewidth]{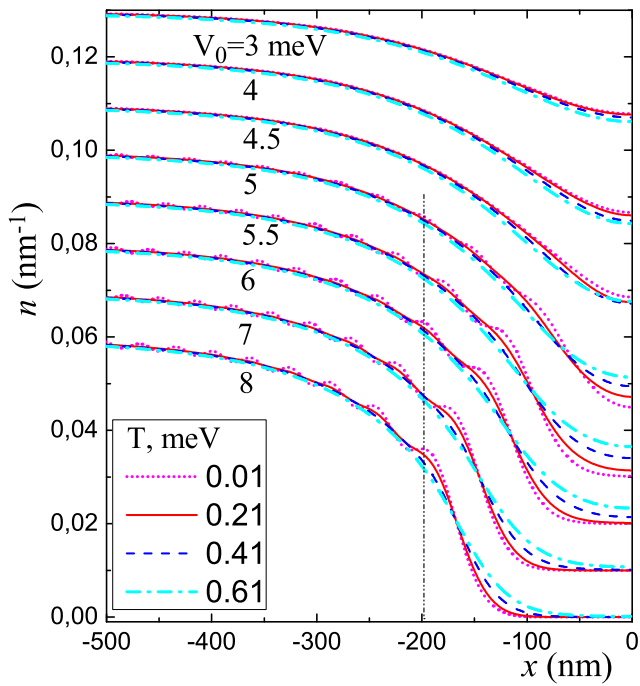}}
\caption{\label{fig1}
1D-electron density calculated with formula~\eqref{eq-4n} at
$E_F=5$~meV for potential $U_0(x)=V_0/\cosh^2(x/a)$, where
the half-width a is fixed $a=200$~nm. Curves for different
$V_0$ are offset by 0.01~nm$^{-1}$ for clarity.
}\end{figure}

To find this correction perturbatively, we first compute the complete
set of wavefunctions for the bare smooth barrier $U_0(x)$ and find the electron
density $n(x)$ at a given temperature:
\begin{equation}\label{eq-4n}
n=\frac{1}{2\pi b} \int_{0}^{\infty}\frac{dE}{(EE_0)^{1/2}} \frac{|\psi_L(x,E)|^2+|\psi_R(x,E)|^2}{1+e^{(E-E_F)/k_BT}},
\end{equation}
where $E_0=\hbar^2/2m^{*}b^2$, $b=1$~nm is the length scale, $m^*=0.067m_e$
is the electron effective mass, and $\psi_L(x,E)$, $\psi_R(x,E)$
is the wavefunction of electrons
incident on the barrier from the left (right). The
amplitude of the incident wave is set to unity. The bare
potential is specified as $U_0(x)=V_0/\cosh^2(x/a)$.
This form is quite appropriate for simulation of short ballistic
channels, including escape to two-dimensional
reservoirs.\cite{Buttiker,pyshkin} This potential does not yield any
features in the transmittance $D(E)$ except the steps of unit
height.\cite{LL} The values of the parameters were taken to
be typical for ballistic quantum wires in a
GaAs/AlGaAs two-dimensional electron gas. The calculated
dependence $n(x,T)$ is presented in Fig.~\ref{fig1}.
%%%%%%%%%%%%%%%%%%%%%%%%%%%%%%%%%%%%%%%
%%%%%%%%%%%%%%%%%%%%%%%%%%%%%%%%%%%%%%%One can
%%%%%%%%%%%%%%%%%%%%%%%%%%%%%%%%%%%%%%%%%%%%%%
\begin{figure}[t]
\centerline{\includegraphics*[width=\linewidth]{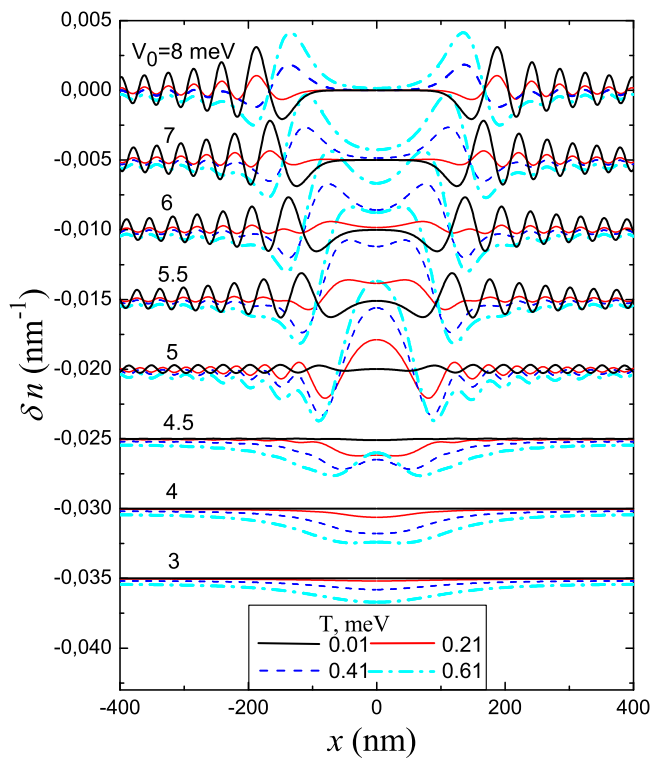}}
\caption{\label{fig2}
Electron density correction $\delta n=n(x)-n_0(x)$
at the same parameters as in Fig.~\ref{fig1}.
Curves for different
$V_0$ are offset by 0.005~nm$^{-1}$ for clarity.
}\end{figure}
%%%%%%%%%%%%%%%%%%%%%%%%%%%%%%%%%%%%%%%%%%%%%%%
One can see that the density strongly changes with increasing temperature
in the transition from the tunnel regime to the open
one. At the lowest temperature there are Friedel oscillations
(FOs) in the tunnel regime, while in the open regime they are
suppressed. Calculated correction $\delta n(x,T)$ is a wide
perturbation of density across the whole barrier
(Figs.~\ref{fig1} and \ref{fig2}).
At $V_0>E_F$
one can see thermally activated increase $n(x,T)$ at the barrier
top; this temperature behavior inverts in the open regime
$V_0<E_F$.
Details of this temperature behaviour are best seen in
$\delta n=n(x)-n_0(x)$, where $n_0(x=0)=n_c(T=0)$,
and  $n_0(x\ne0)=\langle n(x,T=0)\rangle$~is the density averaged
over the Friedel oscillations (Fig.~\ref{fig2}).

\begin{figure}[b]
\centerline{\includegraphics*[width=0.98\linewidth]{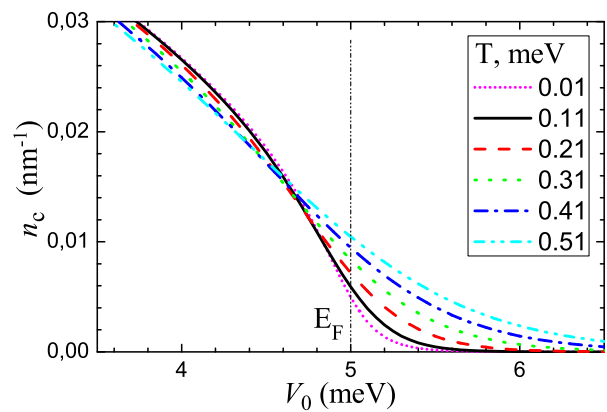}}
\caption{\label{fig3}
Electron density $n_c$ in the center of the barrier versus
the height $V_0$ of the barrier $U_0(x)=V_0/\cosh^2(x/a)$
calculated according to Eq.~\eqref{eq-4n} with $a=200$~nm and $E_F=5$~meV.
}\end{figure}

The most interesting for the analysis of the consequences
of Eqs.~\eqref{eq-2n}, \eqref{deltaU} is the dependence of the
electron density $n_c$ in the center of the barrier on the
bare height $V_0$ at various $T$. Figure~\ref{fig3} shows
quite clearly the details of the $T$-dependent behavior
of $n_c$ and $\delta n_c$, when $U_0(x)$ is an independent variable.
%%%%%%%%%%%%%%%%%%%%%%%%%%%%%%%%%%%%%%%%%%%%%%%%%%%%%%%%%
%%%%%%%%%%%%%%%%%%%%%%%%%%%%%%%%%%%%%%%%%%%%%%%%%%%%%%%%%%%
In the experiment, on the contrary, the gate voltage $V_g$
is varied independently and, according to the electrostatic
calculation (see Appendix \ref{appB} and Refs.~\onlinecite{pyshkin}, \onlinecite{liang}),
there is a linear relation between $n_c$ and
$V_g$ above some small $n_{c0}$ value, so that the temperature
dependence of $n_c(V_g)$ can be neglected. According to
Fig.~\ref{fig3}, this contradiction is resolved under the
assumption that the quantity $V_0$ actually depends on $T$
at constant $n_c$ or $V_g$. The relation between $V_g$ and the
height of the bare barrier $V_0$ is mediated by the electron
wavefunctions and the Fermi distribution. This relation and
$n(x,T)$ do not depend, in the first-order perturbation
theory, on interaction. However, they do contribute
into it.

\section{Corrected potential}
\begin{figure}[b]
\centerline{\includegraphics*[width=0.92\linewidth]{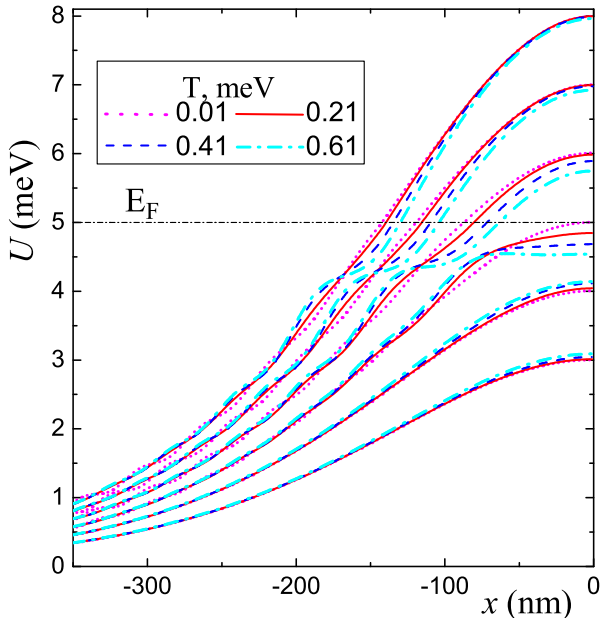}}
\caption{\label{fig4}
Potential $U_0+\delta U(V_0,T)$ calculated from Eq.~\eqref{deltaU-matveev}
with $\alpha=0.2$ for the case of $U_0(x)=V_0/\cosh^2(x/a)$,
$a=200$~nm and $V_0=3,4,5,6,7,8$~meV.
}\end{figure}

\begin{figure}[t]%[htp]
\centerline{\includegraphics*[width=.96\linewidth]{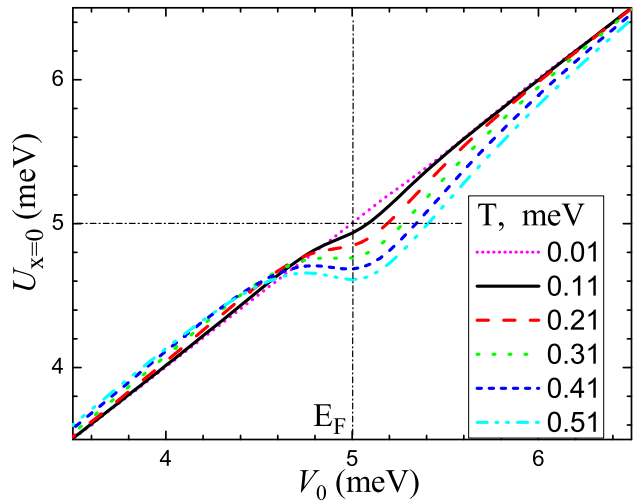}}
\caption{\label{fig5}
Corrected barrier height versus the original barrier
height $V_0$ at the same parameters as in Fig.~\ref{fig4}.
}\end{figure}

According to Eq.\eqref{deltaU} and similar to Ref.~\onlinecite{Matveev94}
the $T$-dependent part
of the interaction-induced
correction to the bare potential $U_0(x)$ was calculated
with the use of the phenomenological formula:
\begin{equation}\label{deltaU-matveev}
\delta U(x)=-\alpha \pi\hbar v_F \delta n(x,E_F,T),
\end{equation}
where $\alpha=\textrm{const}>0$,
$(\hbar v_F)^{-1}$~is the one-dimensional
density of states far from the barrier,
$\delta n=n(x)-n_0(x)$ where $n_0(x=0)=n_c(T=0)$,
and  $n_0(x\ne0)=\langle n(x,T=0)\rangle$~is the density averaged
over the Friedel oscillations.
The interaction-corrected potential $U(x)$ at various
$V_0$ and $T$ values is
shown in Fig.~\ref{fig4}.
%%%%%%%%%%%%%%%%%%%%%%%%%%%%%%%%%%%%%%%%%%%%%%
%%%%%%%%%%%%%%%%%%%%%%%%%%%%%%%%%%%%%%%%%%%%%%%
At high $V_0$ values and low $T$ values,
penetration of an incident electron to the classically
forbidden region of the barrier is very low and the thermal
perturbation of the electron density inside the barrier
is negligible. Therefore, the potential barrier
remains almost unchanged. At $V_0=E_F$, the height of
the barrier $U(x)$ is lowered considerably with an
increase in temperature. At constant $V_0 < E_F$, the barrier
$U(x)$ is raised with $T$, in contrast to the case of
$V_0\geq E_F$. The behavior of the barrier height $U$ is shown
in more detail in Fig.~\ref{fig5}.
For $T=0$ we have $U=V_0$, because $\delta n(x=0)=0$ by definition.
One can see that $U$ becomes independent
%%%%%%%%%%%%%%%%%%%%%%%%%%%%%%%%%%%%%%%%%%%%%%%%%%%%%
%%%%%%%%%%%%%%%%%%%%%%%%%%%%%%%%%%%%%%%%%%%%%%%%%%%
of $V_0$ near $E_F$ at $T>0.1$~meV. There appears a plateau
below
$E_F$. It becomes broader and deeper with an increase in
$T$. The relative correction $\Delta U(x)/V_0$ at the parameters
specified in the figure caption reaches 10\%. This is
close to the limiting value for the present approximation,
which implies that the correction is small. This
limits the growth of $T$ and $\alpha$ in the model. Figure~\ref{fig5} in
combination with Eq.\eqref{eq-2n} provides a qualitative understanding
of the development of the 0.7 conductance
anomaly and the thermopower plateau with an
increase in temperature.
As is seen, the height $U$ of the
temperature-dependent barrier near $E_F$ is stabilized at
about $T$ below $E_F$.
\section{Calculated transport anomalies}
\begin{figure}[t]
\centerline{\includegraphics*[width=0.9\linewidth]{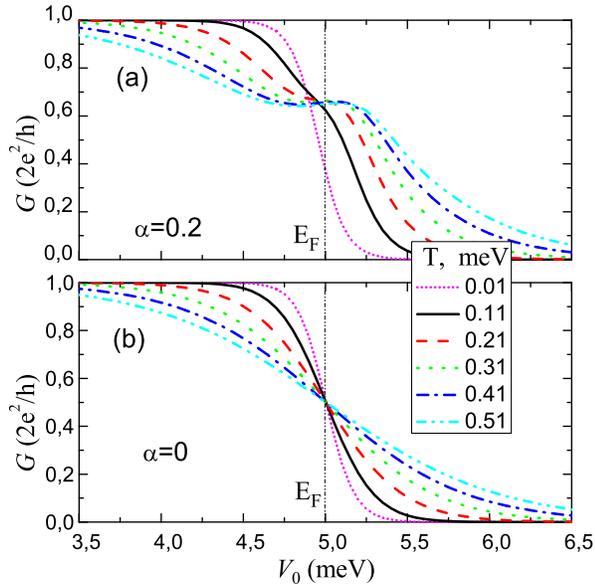}}
\caption{\label{fig6}
(a) Conductance $G(V_0,T)$ calculated for corrected potential
$U(x)=U_0(x)+\delta U(x)$ with $U_0(x)=V_0/\cosh^2(x/a)$, $a=200$~nm
and interaction parameter $\alpha=0.2$. (b) $G(V_0,T)$ for
$U(x)=U_0(x)$
}\end{figure}

\begin{figure}[b]
\centerline{\includegraphics*[width=\linewidth]{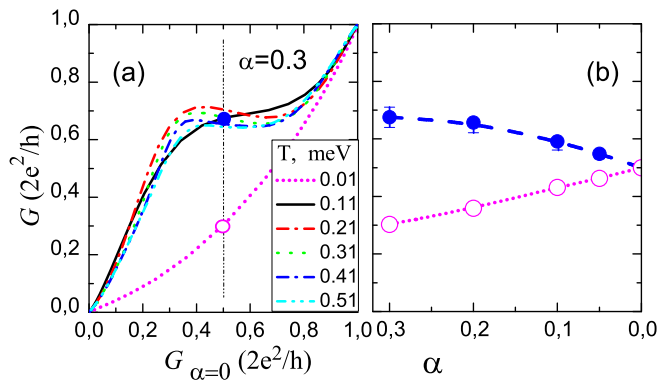}}
\caption{\label{fig7}
(a) Calculated
$G(G_{\alpha=0},T)$ for the same bare potentials as in Fig.~\ref{fig6}a,
but for $\alpha=0.3$.
(b) Calculated
$G(\alpha)$ at $V_0=E_F=5$~meV
for the same bare potentials and different $T$: $0.01\leq T\leq 0.51$~meV.
The upper curve in (b) represents all the curves for $T=0.11\div0.51$,
as they fit within the indicated error bars.
}\end{figure}

Conductance as a function of the barrier height $V_0$ was
calculated with the aid of formulas~\eqref{eq-1n}, \eqref{eq-4n}, \eqref{deltaU-matveev}.
The result
is shown in Fig.~\ref{fig6}. There is an usual conductance step
%%%%%%%%%%%%%%%%%%%%%%%%%%%%%%%%%%%%%%%
%%%%%%%%%%%%%%%%%%%%%%%%%%%%%%%%%%%%%%%
with unit height at $T = 0.01$~meV. However with increasing
temperature additional 0.7-plateau is developed at
$V_0 = E_F$. The width of these plateaux well corresponds
to temperature. Notice that the height of corrected barrier
is almost not changed at the lowest temperatures and only
the distant Friedel oscillations can have influence on
transmission. Indeed, similar to Ref.~\onlinecite{Matveev94},
scattering off Friedel oscillations leads to a small decrease
in transmission coefficient at low but finite temperatures and
a shift of the conductance step to the lower values of $V_0$.
For comparison we show in Fig.~\ref{fig6}b the curves $G(V_0,T)$ calculated for
the bare potential $U_0(x)=V_0/\cosh^2(x/a)$.
Figure~\ref{fig7} shows that conductance behavior is somewhat universal
for elevated $T$ and $\alpha$, and it strongly differs from that for $T\to0$.
Conductance $G(T,\alpha)$ for corrected potential is plotted as a function of $G(T,{\alpha=0})$
 in Figure~\ref{fig7}a.
These conductances are related to each other via common parameter $V_0$.
Conductance $G$ calculated at $E_F=V_0$ as a function of interaction parameter $\alpha$
shows that the height of the $0.7$-plateau is saturated with increasing $\alpha$
(Fig.~\ref{fig7}b).
%%%%%%%%%%%%%%%%%%%%%%%%%%%%%%%%%%%%%%%
%%%%%%%%%%%%%%%%%%%%%%%%%%%%%%%%%%%%%%%

\begin{figure}[t]
\centerline{\includegraphics*[width=\linewidth]{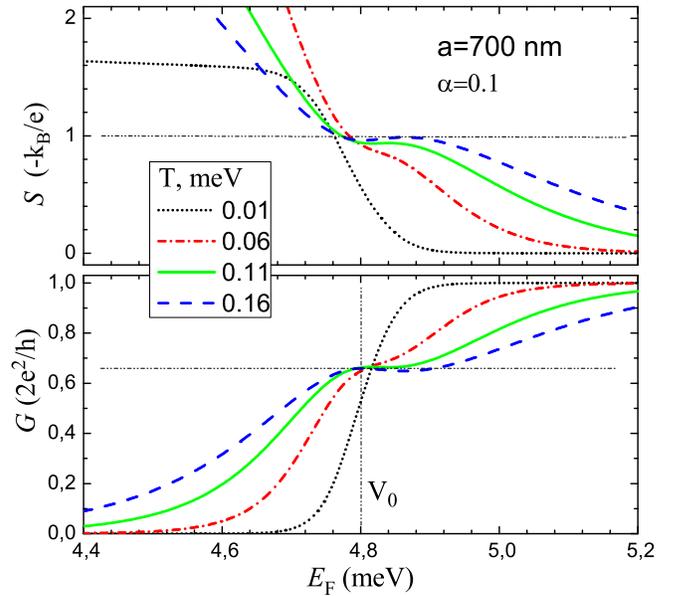}}
\caption{\label{fig8}
Calculated thermopower and conductance of the
one-dimensional channel versus the Fermi energy $E_F$ at
constant $V_0$ (the parameters are indicated in the figure).
}\end{figure}

The dependence of the Seebeck coefficient on $E_F$
and $T$ (Fig.~\ref{fig8}) was computed from Eqs.~\eqref{eq-4n}, \eqref{deltaU-matveev}, and
\eqref{eq-1n}. In this case, we used a larger barrier half-width
and a lower $\alpha$ value than before.
%%%%%%%%%%%%%%%%%%%%%%%%%%%%%%%%%
%%%%%%%%%%%%%%%%%%%%%%%%%%%%%%%
In agreement with the
analysis within approximation~\eqref{eq-2n}
and similar to the
experiment,\cite{Appleyard} $S(E_F)$ exhibits an anomalous step
with a height of 0.9-1.0 of $-k_B/e$. As is clearly seen in
Fig.~\ref{fig8}, this step is formed with an increase in $T$ simultaneously
with the 0.7 conductance anomaly. It is
noteworthy that such a combined evolution with temperature
has not yet been observed. Thus, we propose to make the
respective experiment for the additional proof of the proposed model.

\begin{figure}[t]%[b]
\centerline{\includegraphics*[width=.94\linewidth]{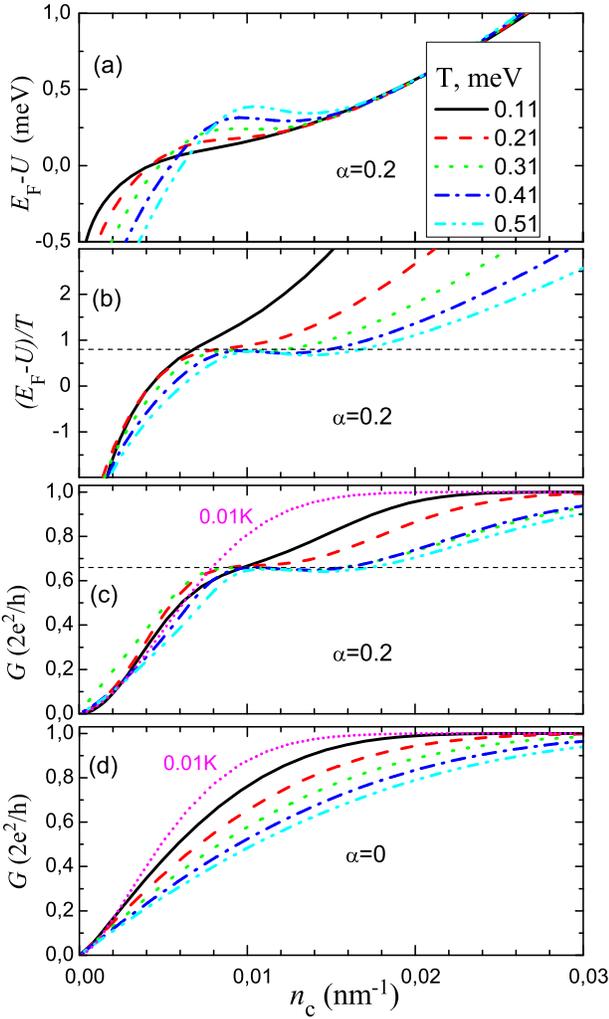}}
%\centerline{\includegraphics*[width=.92\linewidth]{alpha0}}
\caption{\label{fig9}
(a,b) Calculated dependences $E_F-U$, $\eta(n_c)=(E_F-U)/k_BT$
for $E_F=5$~meV, the interaction parameter
$\alpha=0.2$, and the bare potential $U_0(x)=V_0/\cosh^2(x/a)$
with $a=200$~nm. (c,d) Calculated conductance of the
one-dimensional channel with the corrected (b) and the bare (c)
potential versus $n_c$.
}\end{figure}

Though dependences $U(V_0)$, $G(V_0)$, $G(E_F)$ and $G(G_{\alpha=0})$
are easy to calculate, they can hardly be measured.
However, we can compare with the experiment
the respective dependences on the electron density $n_c$
in the center of the barrier, see Fig.~\ref{fig9}.

In the figure, dependences (a)--(c) show the \emph{plateaux} which
appear and become more pronounced with increase of the temperature.
%%%%%%%%%%%%%%%%%%%%%%%%%%%%%%%%%%%%%%%%%%%%%%%%%
%%%%%%%%%%%%%%%%%%%%%%%%%%%%%%%%%%%%%%%%%%%%%%%%%%%%
The shape of the computed plateaux is almost the same as that of
experimental ones (see Figs.~\ref{fig11} and \ref{fig12} in Appendix \ref{appA}).
The widths of the plateaux in Fig.~\ref{fig9}, $\Delta n_c\approx 0.01$~nm$^{-1}$ and
$\Delta V_g\approx0.01-0.02$~V,
agree in several experiments (Figs.~\ref{fig10}--\ref{fig12}), within small
variations of the gate capacitance.
Appendix \ref{appB} discusses the variations in more details.
If correction Eq.\eqref{deltaU-matveev} is zero ($\alpha=0$),
then calculated dependence $G(V_0(n_c,T))$ is almost the same as for
self-consistent potential obtained in 3D-electrostatic potential calculation
(Fig.~\ref{fig13}d).

We made a
number of simplifying assumptions in our 1D-model.
Therefore, a detailed fit of the experimental data to the
calculated curves is hardly appropriate. For example,
it was checked that approximation~\eqref{eq-2n} in the calculation
of the conductance replaces quite well more general formula~\eqref{eq-1n}
(except the case of ultimately low $k_BT$ values).
Thus, the discovered effect is unrelated to the
details of the barrier profile $U(x)$ and is induced merely
by the dependence of the difference $E_F-U$ on $n_c$; i.e., the
correction $\delta U(T)$ in the center of the barrier, where
Eq.~\eqref{deltaU} presumably holds, yielding $\alpha>0$, is crucial. It
is difficult to find $\alpha$ from theoretical considerations.
However, similarity of the experimental and calculated
curves persists under a 50\% variation of $\alpha$ (Fig.~\ref{fig7}).
It is noteworthy
that the effective value $\alpha=0.2$ corresponds to $r^*n_c^*\approx 1.5$;
i.e., the distance $r$ between the probe
electron in the center of the barrier and the exchange-correlation
hole is $1.5\gamma/n_c$. On the basis of the typical
values $n_c\sim0.01$~nm$^{-1}$ (see Figs.~\ref{fig2} and \ref{fig5}), we can
conclude that the conditions used to Eq.~\eqref{deltaU} and Eq.~\eqref{deltaU-matveev} are fulfilled.

\section{Conclusion}

A simple model of anomalous plateaux
in the conductance and thermopower of one-dimensional
ballistic quantum wires has been proposed on
the basis of the Landauer approach with spin degeneracy.
The key points of the model are pinning of the
effective one-dimensional barrier height $U$ at a depth of
$k_BT$ below the Fermi level under a change in the one-dimensional
density in the center of the barrier or the
gate voltage and the inclusion of all (local and nonlocal)
temperature-dependent interaction-induced corrections
via phenomenological formula~\eqref{deltaU-matveev}.

\section*{Acknowledgements}

This work was supported by the Presidium of the
Russian Academy of Sciences, program no. 24, and
the Siberian Branch, Russian Academy of Sciences,
project no. IP130. We are grateful to Z.D. Kvon,
M.V. Budantsev, A.P. Dmitriev, I.V. Gornyi for
fruitful discussions,
and  A. Safonov for translation.

\appendix
%%%%%%%%%%%%%%%%%%%%%%%%%%%%%%%%%%%%%%%%%%%%%%
\begin{figure}[b]
\centerline{\includegraphics*[width=.9\linewidth]{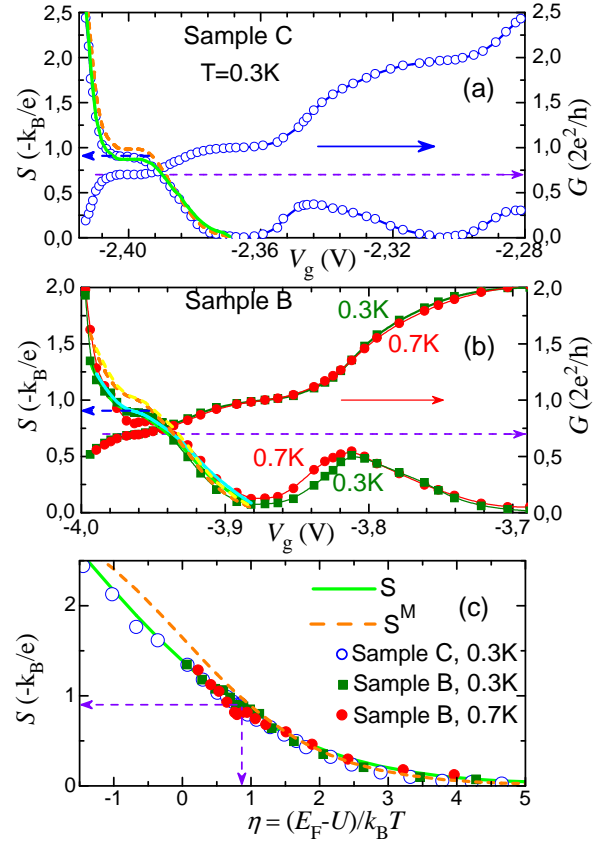}}
\caption{\label{fig10}
Data processing of the gate voltage dependences
of conductance and thermopower from Ref.~\onlinecite{Appleyard} for two
samples with the same geometry of metal gates and two temperatures.
Measured $S$-values are
denoted by open blue and fill red circles, or fill olive square.
Seebeck coefficient $S$ obtained from conductance is
shown by green and cyan line, $S^M$ obtained from $G$ is
shown by orange and yellow line (a,b).
Curves $S(\eta)$ and $S^M(\eta)$ calculated by Eq.\eqref{eq-2n} and
measured $S$-values as a function of $\eta$-values obtained from measured $G$
are shown in panel (c).
}\end{figure}
%%%%%%%%%%%%%%%%%%%%%%%%%%%%%%%%%%%%%%%
\section{Data processing}\label{appA}

We tested the validity of Landauer formulas by the following way.
Combining formulas in approximation~\eqref{eq-2n} it is easy to write
$S=-[(1-G)\ln(1-G)+G\ln G]/G$, $S^M=(\pi^2/3)(1-G)$ where thermopower $S$ and conductance $G$
are measured in units of $-k_B/e$ and $2e^2/h$, respectively.
Thus we can find thermopower from conductance data and compare it to the measured thermopower.
We know about only one paper,\cite{Appleyard} which reports anomalous plateaux for conductance and thermopower
simultaneously. We extracted $S$ and $S^M$ from $G(V_g)$
and ploted the reconstructed and measured points $S(V_g)$ (scaled to common
unit $-k_B/e$). The values corresponding to the first subband almost coincided
(Fig.~\ref{fig10}a,b). One can see that the height of the anomalous conductance
plateau equals 0.7, and the corresponding height of the
thermopower plateau is the same for two samples and agrees
closely with values $S\approx0.8-0.9$, $S^M\approx1$.
Above the first subband the thermopower behaves in accordance
with the Mott law and with the calculations of the peak heights
between zero plateaux.\cite{Proetto,Lunde}
In Refs.~\onlinecite{Proetto} and \onlinecite{Lunde} the height of the first peak
was shown to be approximately equal to $-0.5k_B/e$
if the conductance quantization plateaux are smoothed out, or less than this value if
the plateaux are pronounced.
On the other hand, experimental data in Ref.~\onlinecite{Appleyard} was normalized to
the height of the first peak.
To deal with this, we reduced the measured values of $S$ from
Ref.~\onlinecite{Appleyard} by 2 times to plot the curves in the units of $-k_B/e$.

%%%%%%%%%%%%%%%%%%%%%%%%%%%%%%%%%%%%%%
\begin{figure}[b]
\centerline{\includegraphics*[width=.95\linewidth]{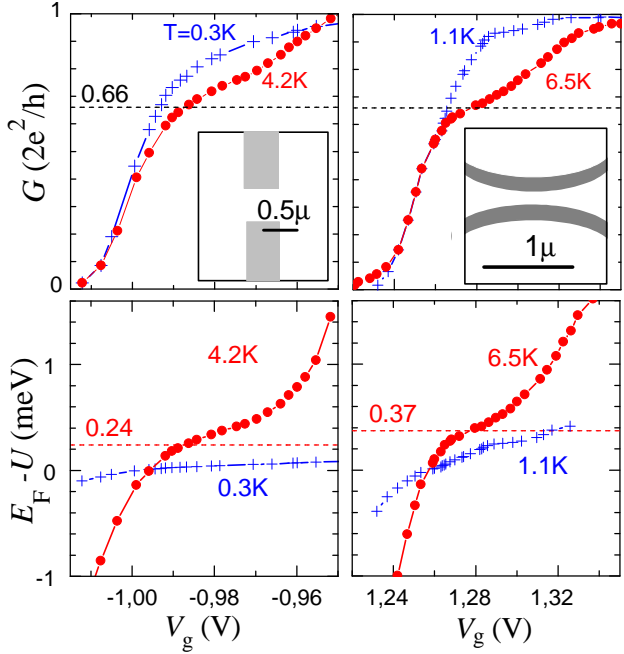}}
\caption{\label{fig11}
Data processing of the gate voltage dependences
of conductance from Refs.~\onlinecite{Liu}, \onlinecite{Kristensen}.
Top: Measured conductance for structures
with split metal gate\cite{Liu} (left) and
in-plane side gate\cite{Kristensen} (right).
The insets show the shape of the gates and etching strips (gray regions) forming
the channel in 2DEG.
Bottom: dependences $E_F-U(V_g)$ extracted from the measured $G(V_g)$
with use of Eq.~\eqref{eq-2n}.
The location of the anomalies in $G$ and $E_F-U$ is
indicated with the dotted lines.
}\end{figure}
%%%%%%%%%%%%%%%%%%%%%%%%%%%%%%%%%%%%%%%
%%%%%%%%%%%%%%%%%%%%%%%%%%%%%%%%%%%%%%
\begin{figure}[b]
\centerline{\includegraphics*[width=0.99\linewidth]{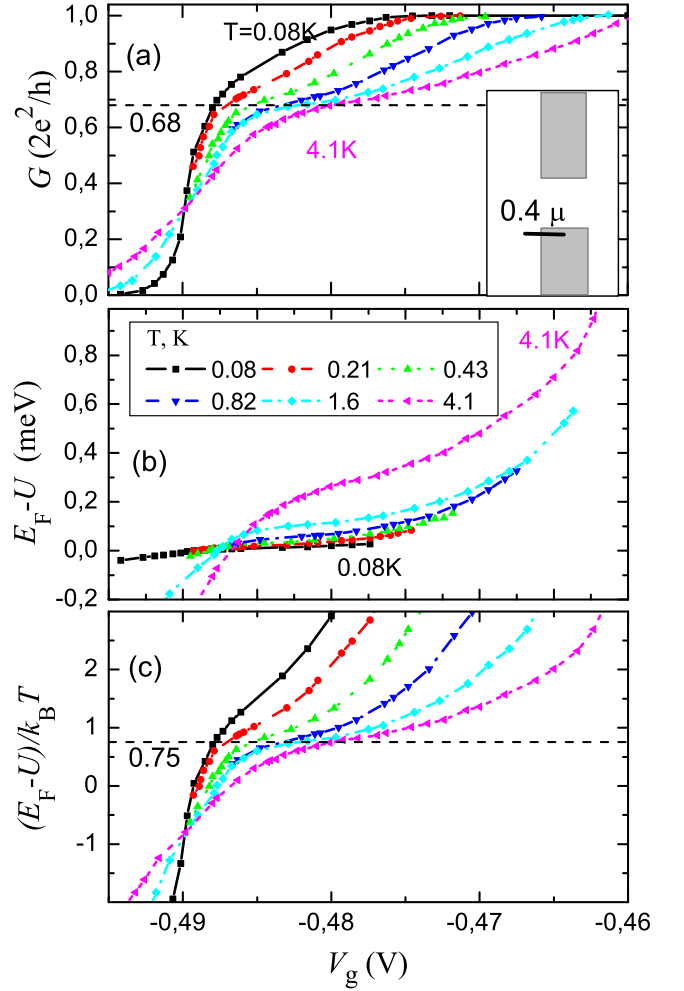}}
\caption{\label{fig12}
Data processing of the gate voltage dependences
of conductance from Ref.~\onlinecite{Cronenwett}.
(a)~Measured conductance for a split metal gate structure.
(b,c)~Dependences $E_F-U(V_g)$ and $\eta=(E_F-U(V_g))/k_BT$
extracted from the measured $G(V_g)$ with help of Eq.~\eqref{eq-2n}.
The dotted lines indicate the position of anomalies in $G$ and $\eta$.
}\end{figure}
%%%%%%%%%%%%%%%%%%%%%%%%%%%%%%%%%%%%%%%

For additional verification of the described interpretation of the experimental data
we extracted $\eta$-values from Eq.~\eqref{eq-2n} and
measured $G$-values for the first subband
and plotted the points $(S,\eta)$ along with the calculated curves $S(\eta)$ and $S^M(\eta)$
using formulas~\eqref{eq-2n}.
Figure~\ref{fig10}c shows a good agreement between the experimental data and
universal curves $S(\eta)$, $S^M(\eta)$,
for different temperatures and devices.
Therefore, we show that the plateaux are present in the $V_g$ dependences
but not in the dependences on $E_F-U$.
This means that the usual assumptions about $U$ being independent of $T$ and
$E_F-U$ being linear in $V_g$ do not work when the first subband
begins getting occupied in the microcontact.
Nevertheless, the spin
degeneracy Landauer approach and Mott approximation
remain valid up to the temperature of 1K in the case of Fig.~\ref{fig10},
including anomalous plateaux.

We used Eq.~\eqref{eq-2n}
% within
%the Landauer single-channel approach
to study the behavior of the reflecting barrier $U=E_F+k_BT\ln(1/G(V_g,T)-1)$
in different quantum point contacts (QPC).
Figures \ref{fig11} and \ref{fig12} show that $U(V_g)$ is pinned with increasing temperature.
In addition, there is a strong temperature dependence of the quantity $U(V_g)$,
which determines the transport.
It is usual to assume that
the main temperature dependence of conductance at any fixed $V_g$
is defined by the Fermi distribution, not the reflecting barrier.
In our case this assumption is not valid, though in Fig.~\ref{fig12}b one can notice
common asymptotics $U(V_g)$ at different $T<4$\,K when conductance approaches to
$2e^2/h$.

%%%%%%%%%%%%%%%%%%%%%%%%%%%%%%%%%%%%%%%%%%%%%%
\begin{figure}[htp]
\centerline{\includegraphics*[width=.8\linewidth]{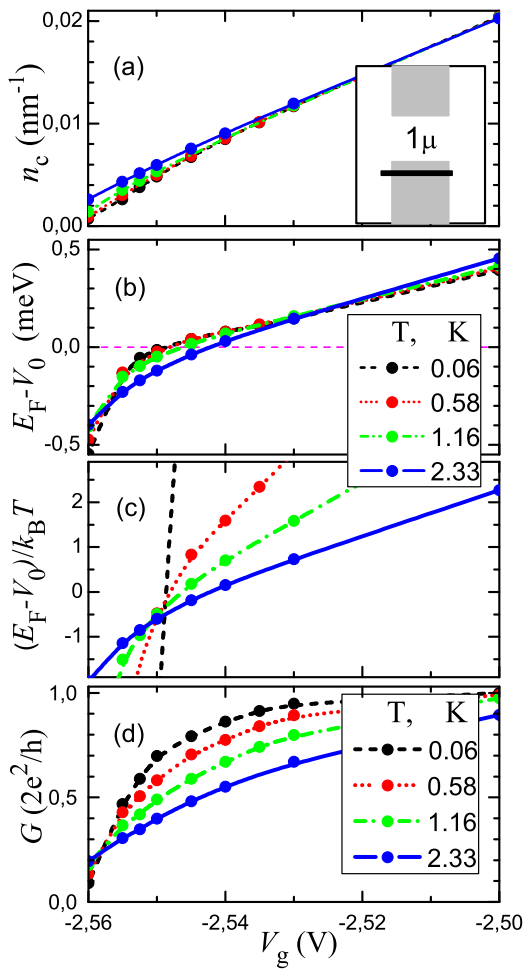}}
\caption{\label{fig13}
(a,b,c) Dependences $n_c(V_g)$, $E_F-V_0(V_g)$ and ($E_F-V_0(V_g))/k_BT$
obtained
from the solution of 3D-electrostatics of the QPC
for usual configuration of split gate and heterostructure
from Ref.~\onlinecite{liang}.
(d) Dependence  $G(V_g)$ obtained
from Eq.~\eqref{eq-1n}
for the lowest one-dimensional subband $V(x)=E_1(x)$, calculated in
3D electrostatic modeling.
}\end{figure}
%%%%%%%%%%%%%%%%%%%%%%%%%%%%%%%%%%%%%%%
\section{Self-consistent calculations}\label{appB}

We computed self-consistency
the 3D-electrostatic potential and the 3D-electron density of a quantum point contact.
We took into account quantization of transverse motion and
exchange-correlation correction of interaction in the local
approximation, determined
by the volume electron density.\cite{hansen,liang}
Fig.~\ref{fig13} shows calculated gate voltage dependences
of the 1D-electron density $n_c$, the first subband bottom
$V_0-E_F$ and $(V_0-E_F)/k_BT$
at the narrowest place of the QPC ($V_0 = E_1(x = 0)$).
Figure \ref{fig13}a shows
that the one-dimensional
electron density $n_c$ in the center of the barrier is almost
independent of $T$ and is linear in $V_g$ starting from small values
$n_{c0}\sim 10^{-3}$~nm$^{-1}$;
i.e., the electric capacitance between the
gate and the quantum wire is conserved at
$G\stackrel{>}{_{\sim}}0.1e^2/h$:
$C_g/e\approx 1/3$~V$^{-1}$nm$^{-1}$.
Notice that for a QPC based on GaAs/AlGaAs heterostructure
capacitance $C_g$ is a weak function of the distance between the gate
and the center of the constriction.
So a typical scale $n_c\approx 0.01$,
corresponding to the width of the anomalous plateau, can be found
from the calculated $C_g$ and the measured interval in $V_g$
($\Delta V_g\approx 0.01\div0.02$~V in Figs.~\ref{fig10}--\ref{fig12}).

It is interesting to compare the behavior of the height of reflecting barrier
$U(V_g)$ (Figs.~\ref{fig11}--\ref{fig12}) discovered by this simple processing of experimental data to
the behavior of a bottom of the first subband $V_0(V_g)$ at the narrowest place,
obtained in the 3D electrostatic self-consistent calculations of potential and electron density
(Fig.~\ref{fig13}b,c).
One may see a qualitative difference between $U(V_g,T)$ (Fig.~\ref{fig12}b,c)
 and $V_0(V_g,T)$ (Fig.~\ref{fig13}b,c).
Calculated dependence $V_0(V_g)$ smoothes with increasing temperature
and there is not any pinning.
Naturally 0.7-anomaly does not appear if formula~\eqref{eq-1n}
is used for potential $V_0(x,T,V_g)$ (Fig.~\ref{fig13}d).
Consequently, ballistic electron feels not self-consistent potential,
but another one (see discussion of formula~\eqref{deltaU} in the main text).

\vfill\eject

\end{document}